\documentclass{jfm}
\usepackage{graphicx}
\usepackage{epstopdf, epsfig}
\usepackage{amsmath}

\author{Stefan Zammert\aff{1,2}
  \corresp{\email{Stefan.Zammert@gmail.com}},
 \and Bruno Eckhardt\aff{2,3}}
 
 \title{Transition to turbulence when the Tollmien-Schlichting and bypass routes coexist}
 
\shorttitle{Tollmien-Schlichting and bypass route to turbulence}
\shortauthor{S. Zammert and B. Eckhardt}

\affiliation{\aff{1} Laboratory for Aero and Hydrodynamics, Delft University of Technology,\\ 2628 CD Delft, The Netherlands\\[\affilskip]
\aff{2} Fachbereich Physik, Philipps-Universit\"at Marburg, D-35032 Marburg, Germany\\[\affilskip]
\aff{3}J.M. Burgerscentrum, Delft University of Technology, 2628 CD Delft, The Netherlands}
\begin{document}

\maketitle

\begin{abstract}
Plane Poiseuille flow, the pressure driven flow between parallel plates, 
shows a route to turbulence connected with a linear instability
to Tollmien-Schlichting (TS) waves, and another one, the bypass transition, 
that is triggered with finite amplitude perturbation. 
We use direct numerical simulations to explore the arrangement of the different 
routes to turbulence among the set of initial conditions. 
For plates that are a distance $2H$ apart and in a domain of width $2\pi H$ and length $2\pi H$
the subcritical instability to TS waves sets in at $Re_{c}=5815$ 
that extends down to $Re_{TS}\approx4884$. The bypass route becomes available above 
$Re_E=459$ with the appearance of three-dimensional finite-amplitude traveling waves.
The bypass transition covers a large set of finite amplitude perturbations.
Below $Re_c$, TS appear for a tiny set of initial conditions that grows with increasing 
Reynolds number. Above $Re_c$ the previously
stable region becomes unstable via TS waves, but a sharp transition to the bypass route can
still be identified. Both routes lead to the same turbulent in the final stage of the transition, 
but on different time scales. Similar phenomena can be expected in other flows where two or more routes to turbulence
compete.
\end{abstract}


\section{Introduction}
The application of ideas from dynamical systems theory to the turbulence transition in flows without linear instability of the
laminar profile, such as pipe flow or plane Couette flow have provided a framework in which many of the
observed phenomena can be rationalized. This includes the sensitive dependence on initial conditions
\citep{Darbyshire95,Schmiegel1997}, the appearance of exact coherent states around which the turbulent state
can form \citep{Nagata1990,Clever1997,Waleffe1998,Faisst2003,Wedin2004,Gibson2009}, 
the transience of the turbulent state \citep{Hof2006, Schneider2008a, Vollmer2009, Kreilos2012}, 
or the complex spatio-temporal dynamics in large 
systems \citep{Bottin1998a,Manneville2009, Barkley2005, Moxey2010, Avila2011a}. Methods to identify
the critical thresholds that have to be crossed before the turbulent state can be reached have been developed \citep{Willis2007,Cherubini2011a} 
and the bifurcation and manifold structures that
explains this behavior in the state space of the system have been identified
\citep{Halcrow2009}. 
Extensions to open external
flows, like asymptotic suction boundary layers \citep{Kreilos2013,Khapko2013,Khapko2013a,Khapko2016} 
and developing boundary layers 
\citep{Cherubini2011,Duguet2012,Wedin2014a} have been proposed.  

Plane Poiseuille flow (PPF), the pressure driven flow between parallel plates, 
shows a transition to turbulence near a Reynolds number of about 1000 \citep{Carlson1982,Lemoult2012,Tuckerman2014}.
In the subcritical range the flow shows much of the transition phenomenology observed in other
subcritical flows, such as plane Couette flow or pipe flow, but it also has a linear instability of 
the laminar profile at a Reynolds number of 5772 \citep{Orszag1971}. 
This raises the question about the relation between the transition via an instability
to the formation of Tollmien-Schlichting (TS) waves and the transition triggered by large amplitude
perturbations that bypass the linear instability 
(henceforth referred to as the "bypass" transition)\citep{Henningson}.
For instance, one could imagine that the exact coherent structures related to the bypass transition
are connected to the TS waves in some kind of subcritical bifurcation. However, the flow structures
are very different, with the exact coherent structures being dominated by downstream vortices 
\citep{Zammert2014b,Zammert2015},
and the TS waves dominated by spanwise vortices. 

In order to explore the arrangement of the different transition pathways we will use direct
numerical simulations to map out the regions of initial conditions that follow one or the other path.
Such explorations of the state space of a flow have been useful in the identification of the sensitive dependence on initial conditions for the transition
\citep{Schmiegel1997,Faisst2004}, and in the exploration of the bifurcations \citep{Kreilos2012,KreilosPRL2014}. 

We start with a description of the system and the bifurcations of the relevant
coherent states in section \ref{sect:coherent}. Afterwards, in section \ref{sect:state space} we describe the exploration of the state space of the
system. Conclusions are summarized in section \ref{sect:conclusion}.

\section{Plane Poiseuille flow and its coherent structures\label{sect:coherent}}

To fix the geometry, let 
 $x$, $y$, and $z$ be the downstream, normal and spanwise directions, 
 and let the flow be bounded by parallel plates at $y=\pm H$. The flow is driven by
 a pressure gradient, giving a parabolic profile for the laminar flow.
Dimensionless units are formed with the height $H$
and the center line velocity $U_0$ so that the unit of time is $H/U_0$ and the
Reynolds number becomes $Re=U_0 H / \nu$, with $\nu$ the fluid viscosity. In these
units the laminar profile becomes $\vec{u}_0=(1-y^2) \vec{e}_x$. 
The equations of motion, the incompressible Navier-Stokes equations,
are solved using \textit{Channelflow} \citep{Gibson2009b}, with a spatial resolution 
of $N_{x}=N_{z}=32$ and $N_{y}=65$ for a domain
of length $2 \pi $ and width $ 2\pi $ and at fixed mass flux.  
The chosen resolution is sufficient to resolve the exact solutions and the transition process
but underresolved in the turbulent case.
In the studied domain, the linear 
instability occurs at $Re_{c}=5815$, slightly higher than the value 
found by \citet{Orszag1971} on account of the slightly different
domain size.

\begin{figure}
\centering
\includegraphics[width=0.5\textwidth]{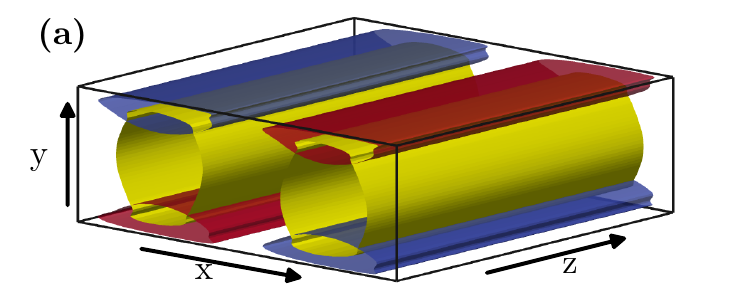}\includegraphics[width=0.5\textwidth]{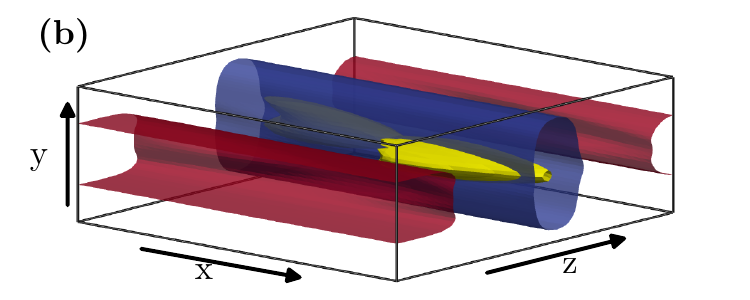}
  \caption{The exact coherent states for the transition to turbulence in plane Poiseuille flow. (a) Visualization of the Tollmin-Schlichting wave $TW_{TS}$. The yellow surface indicates values of $0.3 Q_{max}$ for the Q-vortex criterion. 
The red and the blue surface correspond to $u=\pm 0.1$, respectively. (b) Visualization of the edge state $TW_{E}$ for the bypass transition. As before, the yellow surface indicates values of $0.3 Q_{max}$ for the Q-vortex criterion. The levels for the 
red and blue surfaces are now $u=0.008$ and $u=-0.014$, respectively. \label{fig1}}
\end{figure}

The full velocity field $\vec{U}=\vec{u}_0+\vec{u}$ can be written as a sum of
the laminar flow $\vec{u}_0$ and deviations $\vec{u}=(u,v,w)$.  In the following
we always mean $\vec{u}$ when we refer to the velocity field. 
Tollmien-Schlichting (TS) waves are travelling waves formed by spanwise vortices. 
They appear in a subcritical bifurcation that extends down to $Re\approx 2610$ 
for a streamwise wavenumber of $1.36$.  
The TS wave is independent of spanwise position $z$ and consists of two spanwise vortices, as
shown in figure \ref{fig1}(a).

\begin{figure}
\centering
\includegraphics[]{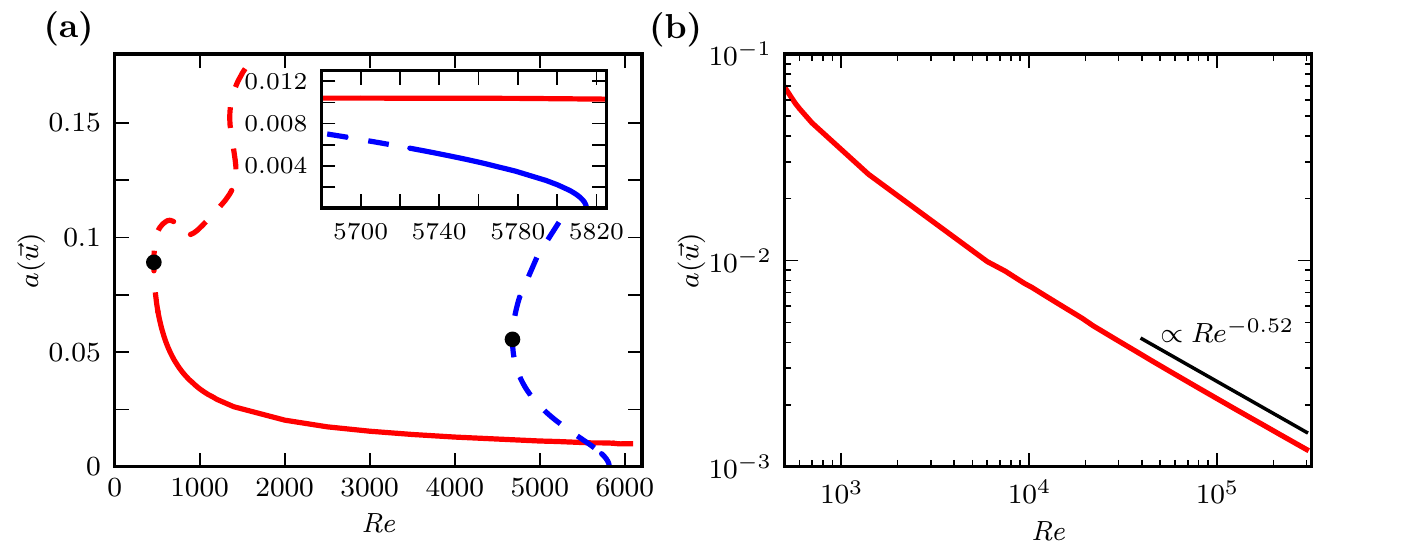}
\caption{The bifurcation diagrams for $TW_{E}$ (red) and $TW_{TS}$ (blue) are shown in panel (a). 
A solid line is used if the travelling wave has just one unstable eigenvalue, 
while a dashed line is used when the wave has further unstable eigenvalues. The inset zooms in on the  
region  where both waves have only one unstable eigenvalues. 
The bifurcation points of the waves are marked with black dots.
In panel (b) the amplitude $a(\vec{u})$ of  $TW_{E}$ is shown in a double-logarithmic 
plot, together with a powerlaw decay like $Re^{-0.52}$ for large $Re$.}
\label{L2NormTWOandTWE}
\end{figure}

The Reynolds number range over which the transition to TS waves is subcritical depends on the 
domain size. For our domain (streamwise wave number of $1.0$)
the turning point is at $Re\approx 4685$. 
A bifurcation diagram of this exact solution, referred to as $TW_{TS}$ 
in the remainder of the  paper, is shown in figure \ref{L2NormTWOandTWE}(a). 
The ordinate in the bifurcation diagram is the amplitude of the flow field
\begin{equation}
a(\vec{u})=||\vec{u}||=\sqrt{\frac{1}{L_{x}L_{y}L_{z}}\int \vec{u}^{2} \ dx \ dy \ dz}.
\end{equation}
A study of the stability of the state in the full three-dimensional space shows that this lower
branch state has only one unstable direction in the used computational domain for $5727<Re<5815=Re_{c}$.
Thus, for these Reynolds numbers the state is an \textit{edge state} whose stable
manifold can divide the state space in two parts \citep{Skufca2006}.
For lower \textit{Re}, there are secondary bifurcations  that add more unstable directions to the
state. Specifically, near the turning point at \textit{Re}$=4690$, the lower branch has acquired 
about 350 unstable directions.
Because of the high critical Reynolds numbers this state cannot explain the transition to 
turbulence observed in experiments at  Reynolds numbers around $1000$ 
\citep{Carlson1982,Nishioka1985,Lemoult2012,Lemoult2013} or even lower \citep{Sano2016}.

The states that are relevant to the bypass transition can be found using the method of edge tracking 
\citep{Toh2003,Schneider2007,Schneider2008}.
Starting from an arbitrary turbulent initial condition, trajectories in the laminar-turbulent-boundary that 
are followed with the edge-tracking algorithm converge to a travelling wave \citep{Zammert2014b} which we referred to as $TW_{E}$ in the following. 
The visualization in figure \ref{fig1}(b) shows that this state has a strong narrow upstream streak, a weaker but more extended downstream streak and streamwise vortices. 
Moreover, $TW_{E}$  has a wall-normal reflection symmetry
\begin{equation}
 s_{y}: [u,v,w](x,y,z)=[u,-v,w](x,-y,z),
\end{equation}
a shift-and-reflect symmetry
\begin{equation}
s_{z}\tau_{x}: [u,v,w](x,y,z)=[u,v,-w](x+0.5 \cdot L_{x},y,-z),
 \end{equation}
and exists for a wide range in Reynolds numbers. 
It is created in a saddle-node bifurcation near $Re \approx 459$ 
(see the bifurcation diagram in figure \ref{L2NormTWOandTWE}(a));
for other combinations of spanwise and streamwise wavelengths
the state appears at a even lower Reynolds numbers of $319$ \citep{Zammert2017}.
 The corresponding lower branch state can be continued to
Reynolds numbers far above $3\cdot 10^{5}$, and its amplitude decreases 
with increasing Reynolds number as shown in 
figure \ref{L2NormTWOandTWE}(b). A fit to the amplitude for large Reynolds numbers gives a scaling like
$Re^{-0.52}$, similar to that of the solution embedded in the edge of plane Couette flow \citep{Itano2013a}.
A stability analysis of the lower branch of $TW_{E}$  shows that the travelling wave has 
one unstable eigenvalue for $510<Re<5850$. Therefore, $TW_{E}$ is a second travelling wave with a stable 
manifold that can  divide the state space into two disconnected parts. How the two edge states interact
and divide up the state space will be discussed in section \ref{sect:state space}.

At $Re=510$  the lower branch undergoes a supercritical 
pitchfork bifurcation that breaks the $s_{y}$ symmetry and adds a second unstable eigenvalues for $Re<510$.
The upper branch of the travelling wave has three unstable eigenvalues for $Re<1000$.  
Investigation of different systems which show subcritical turbulence
revealed that bifurcations of exact solutions connected to the edge state of the system lead to the formation 
of a chaotic saddle that shows transient turbulence with exponential distributed lifetimes \citep{Kreilos2012,Avila2013}.
In the present systems the formation of chaotic saddle 
cannot be studied in detail since it takes place in an unstable subspace.
However, previous investigations in a symmetry restricted system did show that the 
states follow such a sequence of bifurcations to the formation of a chaotic saddle  \citep{Zammert2015,Zammert2017}, so that we expect that also the states in the unstable subspace
follow this phenomenology.

The two travelling waves described above are clearly related to the two different transition mechanism 
that exist in the flow. For Reynolds numbers below the onset of TS waves (here: $Re_c=5815$), 
initial conditions that start close to $TW_{E}$ in the state space will either decay 
or become turbulent without showing any approach to a TS wave: 
they will follow the bypass transition to turbulence. 
Initial condition that start close to $TW_{TS}$ can also either decay or swing up to turbulence, but
they will first form TS waves. Above $Re_c$ all initial conditions will show a transition 
to turbulence, but it will still be possible to distinguish whether they follow the bypass or
TS route to turbulence, as we will see.

\section{State space structure\label{sect:state space}}

In order to explore the arrangement of the different routes to turbulence in the space of initial
conditions we pick initial conditions and integrate them until the flow either becomes turbulent or 
until it returns to the laminar profile. The initial conditions are taken in a two-dimensional slice of the
high-dimensional space, spanned by two flow fields $\vec{u}_{1}$ and $\vec{u}_{2}$.
The choice of the flow fields allows to explore different cross sections of state space. For the most
part, we will use $\vec{u}_1$ and $\vec{u}_2$ to be the travelling waves  $TW_{E}$ and $TW_{TS}$,
so that both states are part of the cross section.  The initial conditions are then 
parametrized by a mixing parameter $\alpha$ and an amplitude $A$, i.e.,
\begin{equation}
\vec{u}(\alpha,A)=
A \frac{ (1-\alpha) \vec{u}_{1} + \alpha \vec{u}_{2}}{\left|\left|(1-\alpha) \vec{u}_{1} + 
\alpha \vec{u}_{2}\right|\right|}.
\end{equation}
For $\alpha=0$ one explores the state space along velocity field $\vec{u}_1$ and for
$\alpha=1$ along velocity field $\vec{u}_2$.
If the upper and lower branch of $TW_{E}$ are used to create such a slice, one recognizes that
the turbulence in PPF appears in similar chaotic bubbles as in 
plane Couette \citep{Kreilos2012,Zammert2016x}.

\begin{figure*}
\centering
\includegraphics[width=0.52\textwidth]{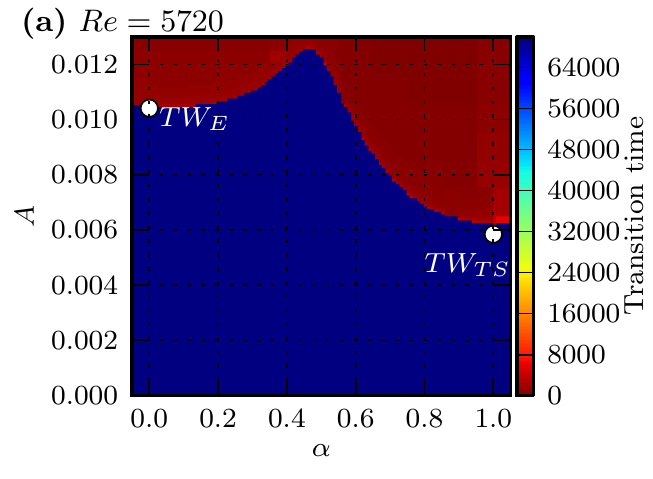}\includegraphics[width=0.52\textwidth]{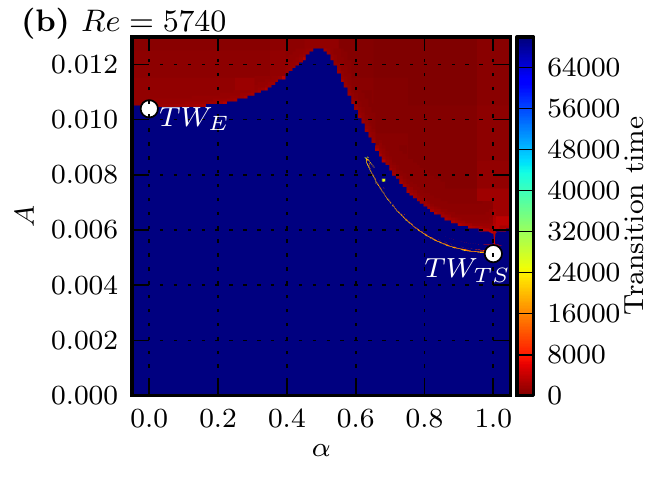}
\\
\includegraphics[width=0.52\textwidth]{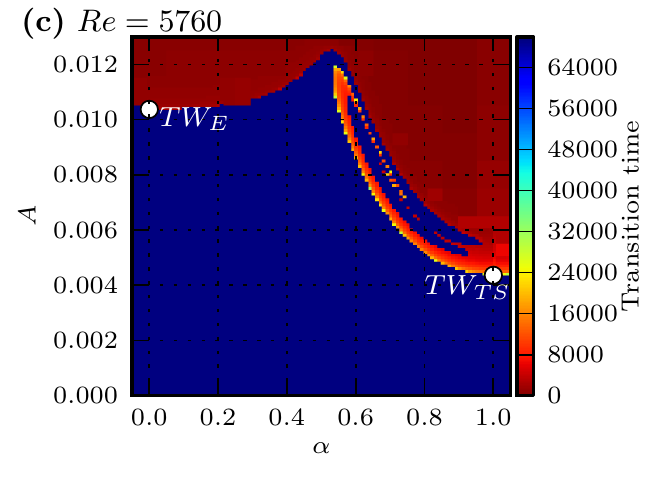}\includegraphics[width=0.52\textwidth]{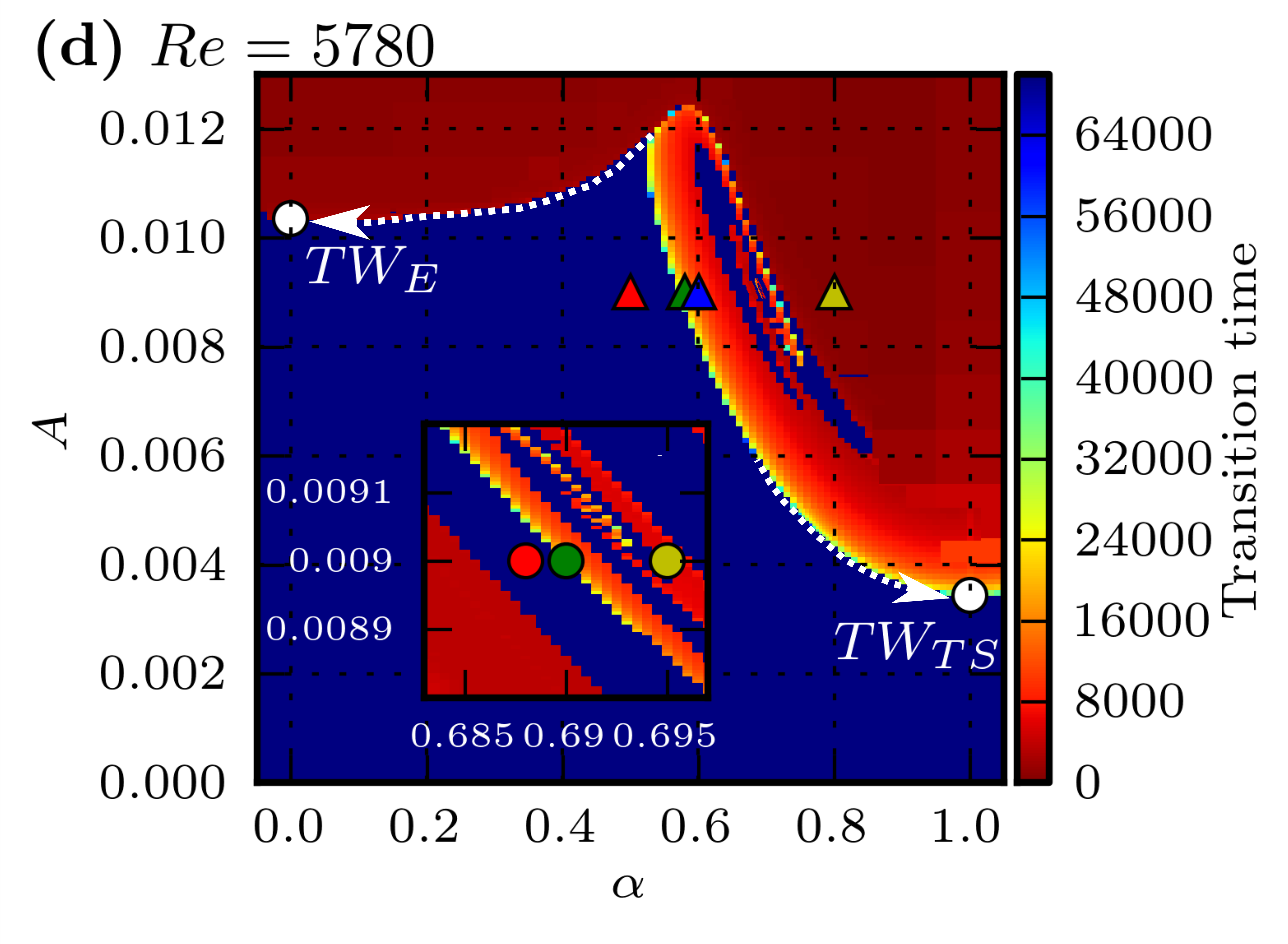}
\\
\includegraphics[width=0.52\textwidth]{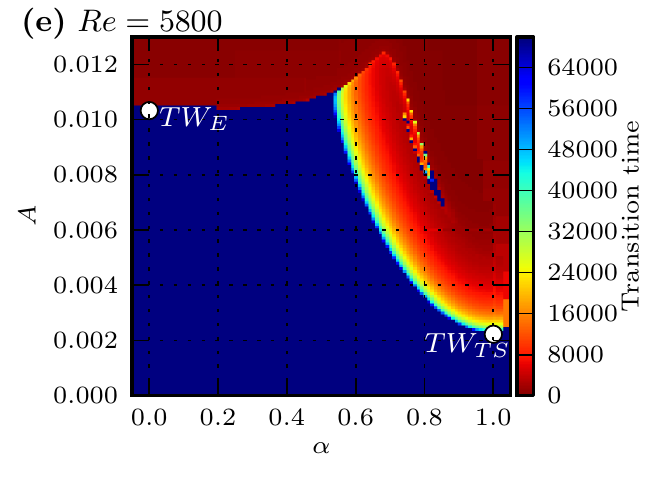}\includegraphics[width=0.52\textwidth]{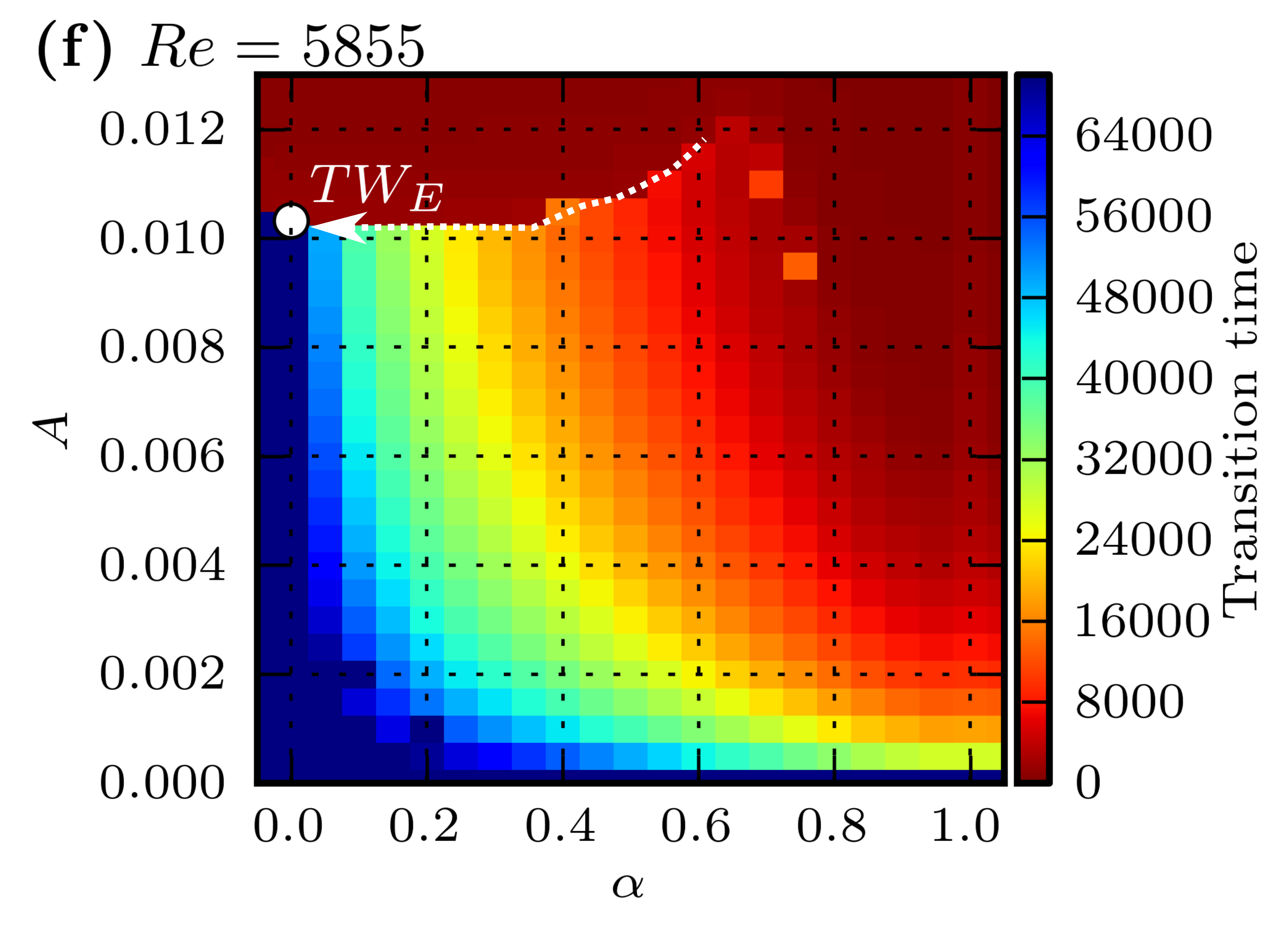}
\caption{Two-dimensional slices of the state space for various Reynolds numbers. In panel (a-e), where
$Re<Re_{c}=5815$, the parameter $\alpha$ interpolates 
between the flow fields of both traveling waves, and both are indicated by white dots in the figures
In panel (f), where  $Re>Re_{c}$ the plane is spanned by the flow field of $TW_{E}$ and by 
the unstable TS-mode of the laminar state. In all panels the color indicates the time it takes to reach the
turbulent states, up to a maximum integration time of $70000$ time units. Accordingly, initial
conditions that do not become turbulent or return to the laminar state are indicated by dark blue.
The dashed lines in panels (d) and (f) indicate the stable manifolds of the corresponding fixed
points. The colored triangles and dots in panel (d) mark initial conditions whose time evolution
is shown in figure \ref{fig_Trajectories}.}
\end{figure*}
Lower branch states are relevant for the transition to turbulence, so begin by exploring the
slice spanned by the lower branches of $TW_{E}$ and $TW_{TS}$. We assign to each 
initial condition the time it takes to become turbulent, with an upper cut-off for initial conditions
that either take longer or that never become turbulent because they return to the laminar profile.
Color coded transition-time plots are shown in figure \ref{fig_TransTimes}(a) - (e) for
different Reynolds numbers below $Re_{c}$.
The boundary between initial conditions that relaminarize and those that become turbulent stands out
clearly. They are formed by the stable manifold of the states and their crossings with the cross section.
Parts of the stable manifold are indicated by the dashed white lines for better visibility. 
The part of the laminar-turbulent boundary connected with $TW_{E}$ can be distinguished from that 
connected to $TW_{TS}$ by the huge differences in transition times:
for $TW_{TS}$ transition times are significantly longer and even exceed $2\cdot 10^{4}$ time units. 
The interaction between the two domains is rather intricate. 
For Reynolds number $5780$, shown in  figure \ref{fig_TransTimes}(d), it seems that the borders 
do not cross but rather wind around each other in a spiral shape down to very small scales.
Although the wave $TW_{TS}$ has still only one unstable eigenvalues, 
the size of the structure that is directly connected to $TW_{TS}$ shrinks 
with decreasing \textit{Re} and is not visible in these kind of 
projection for $Re<5727$ where $TW_{TS}$ has more than one unstable eigenvalue. 
\begin{figure}
\centering
\includegraphics[]{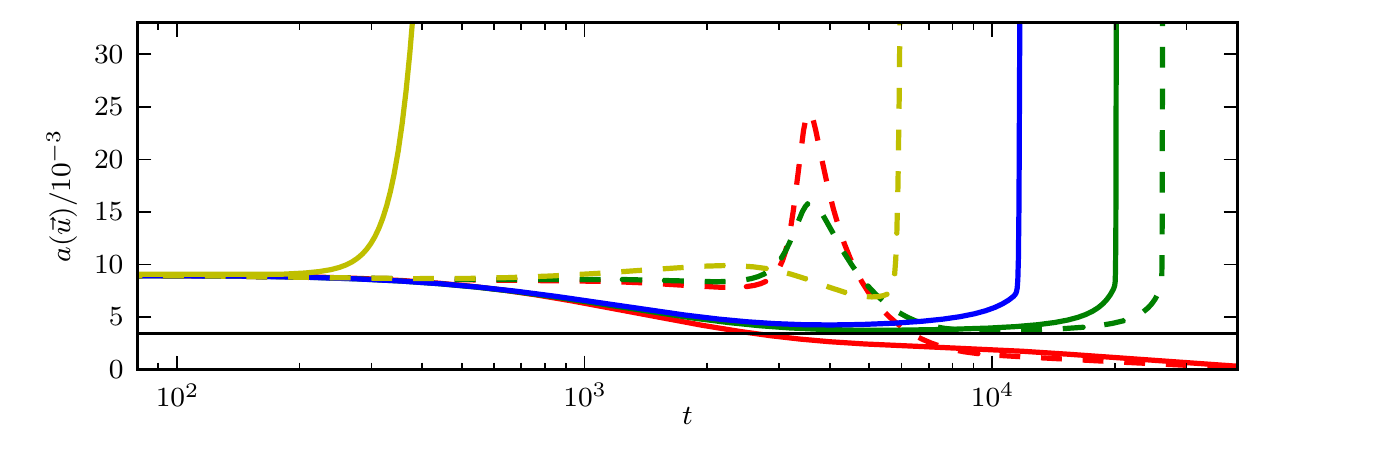}
\caption{Time evolution of the velocity amplitude $a(\vec{u})$ for the initial
 conditions marked in figure \ref{fig_TransTimes}(d) for $Re=5780$.
The initial conditions for the trajectories drawn with solid and dashed lines are marked in figure \ref{fig_TransTimes}(b) with triangles and circles, respectively. 
The black line indicates the amplitude of the lower branch of the TS state 
$TW_{TS}$ at $Re=5780$. Some of the initial conditions that miss the bypass transition
become turbulent nevertheless because they are captured by the TS instability 
(e.g. the full blue, and both green lines).}
\label{fig_Trajectories}
\end{figure} 

In figure \ref{fig_Trajectories} the evolution of the amplitude for different initial conditions marked in figure \ref{fig_TransTimes}(d) 
is shown. The green and blue lines are typical representatives of the slow TS transition. Starting with
a three dimensional initial condition, their amplitude decays and the two-dimensional TS wave $TW_{TS}$, whose amplitude is marked by the black line, is approached.
Afterwards, they depart from $TW_{TS}$ again, which is a slow process because of the small growth rate.
Ultimately, the transition is caused by secondary instabilities of the TS waves \citep{Herbert1988}.

The solid yellow line in figure \ref{fig_Trajectories} is an initial condition that undergoes bypass transition. It
quickly swings up to higher amplitudes and does not approach the TS wave on its way to turbulence.
The dashed yellow line is of an intermediate type. It takes a long time to become turbulent but it does not
come very close to the TS wave. The relation between time-evolution, transient amplification,
and final state is complicated and non-intuitive. For instance, the dashed red and green trajectories 
share a transients increase near $t\approx4000$, but differ in their final state: the red curve, with the higher
maximum, eventually returns to the laminar profile, but the green curve, with the smaller maximum,
approaches the TS level and eventually becomes turbulent following the TS route.
Similarly, the red, blue and green continuous lines start with high amplitude slightly below the 
threshold for the bypass route. They all decay, but while the red initial conditions ends up on 
the decaying side of the TS wave, the green and blue one eventually become turbulent
via the TS route.

For plane Couette flow it was found that a small chaotic saddle can appear inside of existing larger ones \citep{KreilosPRL2014}. 
There, trajectories that escape from the inner saddle are still captured by the outer one. The appearance of TS transition
in PPF follows a comparable mechanism. With increasing Reynolds number the chaotic saddle of subcritical bypass turbulence is surrounded
by the stable manifold of the TS wave that above $Re=5727$ can separate two parts of the state space and therefore prevent trajectories in the interior from becoming laminar.

With increasing Reynolds number the number of initial conditions becoming turbulent increases. 
Finally, for $Re>Re_{c}$ no initial conditions that return to the laminar state exist anymore. 
Nevertheless, also in this supercritical regime a sudden change in the type of transition can be identified:
when the amplitude increases and crosses the stable manifold of $TW_{E}$, the transition time
drops dramatically and turbulence is reached via the bypass route. 
In the state space visualization for $Re=5855$ that is shown in figure \ref{fig_TransTimes}(c), these change of the transition type presents itself in the rapid drop of the transition time with increasing $a$ for $\alpha$ values between $0$ and $0.6$.

In the supercritical range the stable manifold of the bypass edge state $TW_{E}$ separates initial conditions undergoing the quick bypass
transition form initial conditions that become turbulent by TS transition.
The state space picture at a higher Reynolds number of $6000$ looks qualitatively similar to 
the one shown in figure \ref{fig_TransTimes}(c), including the switch from TS to bypass transition when
the  stable manifold of $TW_{E}$ is crossed.

\section{Conclusions \label{sect:conclusion}}

We have explored the coexistence of two types of transition in subcritical plane Poiseuille flow 
connected with the existence of states dominated by streamwise and spanwise vortices
(bypass and TS transition). 
Probing the state space by scanning initial conditions in two-dimensional 
cross sections gave information on the sets of initial conditions that follow one or the other route to turbulence. 
The results show that the transition via TS waves initially occupies a tiny region of state space. As this
region expands it approaches the bypass-dominated regions, but a boundary between the two 
remains visible because of the very different times needed to reach turbulence. This extends to the
parameter range where the laminar profile is unstable to the formation of TS waves. 

The results shown here are obtained for small domains, where the extensive numerical computations
for very many initial conditions are feasible. For larger domains, the corresponding exact coherent
structures are localized, as shown by \citet{Jimenez1990a} and \cite{Mellibovsky2015} for TS waves and 
by \citet{Zammert2014b} for the bypass transition. Since the bifurcation diagrams for the 
localized states are similar to that of the extended states, we anticipate a similar phenomenology 
also for localized perturbations in spatially extended states. 

The methods presented here can also be used to explore the relation between bypass transition 
and TS waves in boundary layers \citep{Duguet2012,Kreilos2016}. More generally,
they can be applied to any kind of transition where two different paths compete: examples include
shear driven or convection driven instabilities in thermal convection \citep{Clever1992,Zammert2016b}, 
the interaction between transitions driven by different symmetries \citep{Faisst2003,Wedin2004,Schneider2008}, 
or the interaction between the established subcritical scenario and the recently discovered
linear instability in Taylor-Couette flow with rotating outer cylinder \citep{Deguchi2017}.


This work was supported in part by the German Research Foundation (DFG) within Forschergruppe 1182.

\bibliographystyle{jfm}

\end{document}